# A new artificial material approach for flat THz frequency lenses

Giorgio Savini,[1,*] Peter A.R. Ade,[2] and Jin Zhang[2]

[1]*Department of Physics & Astronomy, University College London, London WC1E 6BT, UK*
[2]*School of Physics & Astronomy, Cardiff University, Cardiff CF24 3AA, Wales, UK*
[*]*g.savini@ucl.ac.uk*

**Abstract:** Stacked layers of metal meshes embedded in a dielectric substrate are routinely used for providing spectral selection at THz frequencies. Recent work has shown that particular geometries allow the refractive index to be tuned to produce practical artificial materials. Here we show that by spatially grading in the plane of the mesh we can manufacture a Graded Index (GrIn) thin flat lens optimized for use at THz frequencies. Measurements on a prototype lens show we are able to obtain the parabolic profile of a Woods type lens which is dependent only on the mesh parameters. This technique could realize other exotic optical devices.

**1. Introduction**

The development of variable refractive index optics dates back to Maxwell's spherical fish-eye lens. In 1905, R W Wood [1] used a gelatin-dipping technique to produce variable index cylindrical lenses, although he mentions that Schott had already produced variable index glass optics by rapidly cooling molten glass in an iron tube. It is of no surprise that far infrared optics attempts to reproduce similar effects by taking advantage of material-engineering to modify the structure on sub-wavelength scales. In 1967 Ulrich [2] elaborated on a previous work [3] suggesting the use of copper evaporated periodic metal structures in order to obtain a thin surface which would act as a low pass electromagnetic filter for wavelengths greater than the characteristics size of the periodic cell. The field that originated from this found immediate application in many branches of experimental microwave techniques with a particular success story in astronomical instrumentation, where the once single layer of periodic holes was complemented by the negative equivalent in a Babinet fashion. These structures were classified as capacitive or inductive, dependent on their transmission properties in a direct analogy with a transmission line [4]. Capacitive and Inductive grids have since been modeled and their detailed frequency behavior investigated [5,6] with increasing precision [7] and wider parametric studies [8]. Multiple layers were soon proposed [2,6,9,10] to produce the first "engineered materials" in the sub-mm THz region able to control the selective reflection and transmission of far infrared radiation covering wavelength from 10μm to the millimeter region. It is only in recent years that in the wake of the many successes of the photonics community engaging in photonic crystals [11] and in particular sub-wavelength periodic structures used in surface plasmon physics [12,13] that it was realized that the same technology employed for the construction of these compact and durable astronomical filters could be exploited for a different purpose where the properties of the overall macroscopic material were of interest rather than the specific combination of metallic meshes with a given set of patterns and spacings. In analogy to the creation of a given refraction index material via dielectric loading, Zhang et al. [14] showed that a closely stack set of capacitive grids with period "g" and layer distance $d < \lambda$ could create an effective material with a given characteristic impedance (refractive index) with only mild chromatic behavior. As a base dielectric material, polypropylene was chosen because of its low absorption and because of the ease in fusing multiple-layers containing the metalized patterns to form a robust material [15]. Validation was achieved by engineering a material with uniform mesh geometry of refractive index "n" to match the requirement of anti-reflection coating for a slab of sapphire at a given wavelength, the root-square of $n_{sapphire}$. Here we advance this concept by grading the refractive index across the plane orthogonal to the optical axis varying the mesh geometry to create a lens [16]. In this approach we take advantage of two huge resources: the well developed theory of gradient index optics available from decades of successful optics manufacturing and testing at visible wavelengths ([17-19] and references therein) and the recently imaginative and analytically exhaustive transformation optics [20-22]. The simplest approximation for a Wood lens as cited by Fischer [23] is adopted to obtain a flat and thin (2mm) Gradient-Index (GrIn hereafter) lens of d = 7cm in diameter optimized for wavelengths $\lambda \cong 2mm$. We show that this lens works efficiently over a large spectral bandwidth due to the weak chromatic nature of the engineered material, and report its design, build, spectral and spatial performance. While there are a large foreseeable number of applications which will benefit from this technology, we wish to point out that our selection of $\lambda = 2mm$ as the optimized wavelength is driven by the search for a low-weight, easy to coat, cryogenically proven and space-tested technology alternative for large highly-curved polymer lenses for future satellite instrumentation.

## 2. Results

*2.1 Modelling the spatial distribution of "n"*

Two independent approaches were combined in order to design the gradient index lens: a Transmission Line Model (TLM)[7] and a Finite Element Analysis model (HFSS)[24]. The TLM provides the overall lumped impedance framework whilst HFSS provides a reliable map of the parameter space associating index of refraction "n" with number of mesh layers "N", inter layer spacing "t", cell period "g" and cell half-gap "a" according to the same formalism adopted in [2]. Some of these results were published in Zhang2009[14] as part of an analysis aimed at tuning an artificial material refractive index for anti-reflection coatings. The latter two parameters define the metallic layer geometry which dictates the complex impedance of the metal mesh structure [5], as used by the TLM, and the cut-off wavelength as the capacitive mesh acts as a low pass filter. Fig.4 of [14] shows how the variation of the ratio of these parameters a/g, produced an appreciable change in the equivalent refractive index with little wavelength dependence. This study has now been performed in more detail with the equivalent refractive index extracted by two independent methods; the first references the retardation relative to free space whilst the second uses a direct inversion of the Fabry-Perot etalon expression [25] for a parallel plate standing wave generated by both the TLM and the HFSS models for a range of frequencies below the cut-off. The resulting index values plotted in Fig.1 are limited from the low side by the value of the dielectric employed (in this case polypropylene with $n_{PP} = 1.48$) as a substrate for the metal pattern deposition. In the design of the prototype lens there were also practical limitations imposed on the choice of "N" and "t" by the manufacture process.

The dielectric spacers comprise sheets of 10μm thickness so large values of "t" require multiple layers whilst the number "N" of identical metallic patterned layers which are photo-lithographically replicated from a single chrome-on-glass master, require careful processing, assembly and alignment to create the final lens. Thus for the first prototype we chose to use a modest grid-periodicity, g = 200μm, making replication easier, and varied the effective impedance by tapering the "a/g" parameter from 0.06 at the lens centre, corresponding to a refractive index of $n_0 \cong 2.6$, to 0.50 at the outer edge of the metalized region. This edge value is little different to the outmost region where there is no copper so the refractive index effectively tapers to $n_{PP} = 1.48$. This tapering was designed with a parabolic curve as is required by a Wood type lens. The smoothness of the refractive index change is limited by the photo-mask accuracy (step changes from cell to cell) and control of the photo-lithographic replication.

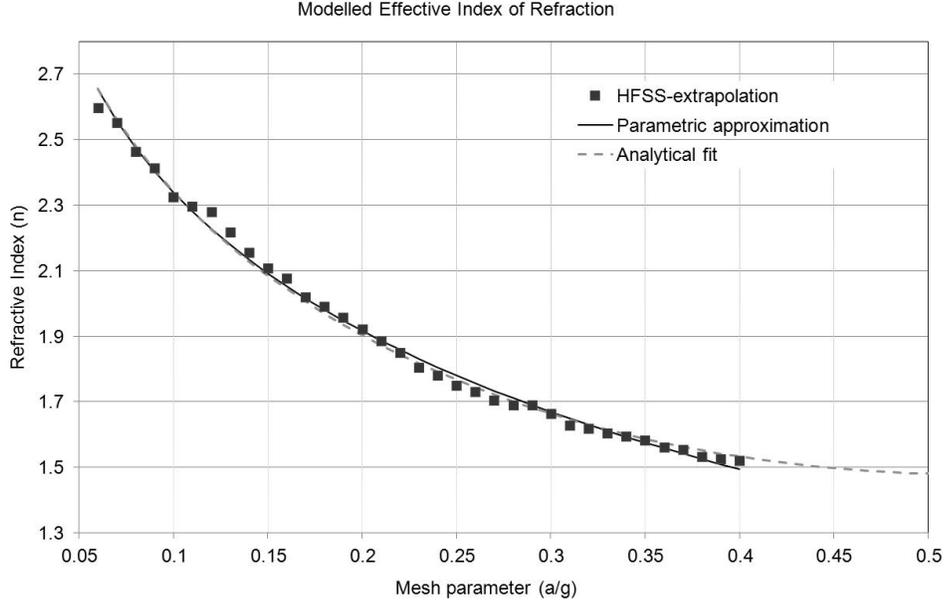

Fig.1. Modeled index of refraction with HFSS, best analytical fit function (dotted line) and parametric approximation adopted in the patterning algorithm used in the master generation as discussed below.

A numerical fit to the simulated parameters was performed to obtain a smooth algorithm for the structure defining taper as shown in Fig.1. The resulting function is approximated by:

$$n(a/g) = -p_1 \ln(a/g) + p_2 \qquad (1)$$

with coefficients $p_1 = 0.61$ and $p_2 = 0.935$. An improvement of the physical model used in the fit can be obtained by using a more complex algorithm to achieve an asymptotic taper into the no-mesh region but this refinement was not considered necessary for the prototype. Note that the minimum value of "n" predicted by this equation diverges for $a/g \rightarrow 0$ as theoretically this value coincides with a fully reflective surface. The maximum model value of $n_{max} \cong 2.7$ is important here as it defines the lens recipe and dictates the low-pass filtering properties of the stack of capacitive meshes.

*2.2 Wood lens recipe.*
For the proof of concept a lens was designed to be similar to standard High Density Poly-Ethylene lenses used in our own mm-wave instrumentation in order to be able to directly compare their performances. The design was for a lens of focal length f = 250mm and radius R = 34mm such that [23]:

$$n = n_0 - \frac{r^2}{2df} \qquad (2)$$

with a central value $n_0 \cong 2.65$ in order for the profile of the lens to terminate at the substrate value 1.48 when reaching the edge of the intended lens of thickness d=2mm.

From Zhang2009 it is shown how a close spacing of a few microns for a g = 100μm allows to obtain an index of refraction $n \cong 5$ at λ = 3mm. Here we are not in search of high index of refraction as much as we need to combine a sufficiently high index at the lens centre $n_0$ with a not-so-fast index variation in order for the condition[17] Δn/n << 1 to hold on scales comparable to the wavelength or more importantly to the scale that we can control which is that of the periodic cell. Such considerations lead to imposing the cell condition on the minimal variation of the index "n" to the first derivative of Eq.2 :

$$\Delta n = n(0) - n(\Delta r) = \frac{(\Delta r)^2}{d\mathrm{f}} \qquad (3)$$

We can quickly show how this is not achieved by our photo-lithographic technique which has an achievable resolution of 1μm. At the centre of our lens this condition translates to a minimum variation of index from Eq.3 of $\Delta n \sim 10^{-5}$ which corresponds to a variation of less than 1 nm on the photo-lithographic mask. However, for the wavelengths involved we are effectively averaging over tens of cells in the mesh structure so the uncertainty on the mesh geometry will average out. At a radius r > 5mm the change in half-gap is now greater than the dimensional tolerance and therefore more precisely reproduces the profile.

The manufacturing technique uses 10μm layers of bi-axial polypropylene to create 100μm thick dielectric spacers interspersed with a manageable number of mesh layers (≤ 20) to create the 2mm thick lens. This stack is then placed in a hot oven to fuse all the layers together making a mechanically robust structure as is used in the manufacture of metal mesh filters [26]. Thus the first prototype required a total of 200 layers of which 20 equally spaced layers are patterned with the copper mesh structure.

### 2.3 Build and measurements

One of the common features of periodic grids and meshes is the ease of manufacture due to the periodicity and ease of programming of such a structure in photo-lithographic masters and beam-etching facilities. A complication here is the requirement to have a spatially varying structure to provide a relatively smooth variation of the parameters involved in the geometry. The photo-master manufacture was achieved by providing a coordinate vector file rather than an image file to the mask generator (similarly to electronic circuits).

A script was compiled to generate the pointers to the squares based on Eq.2 and Eq.3, transforming the problem of mask file size from a graphical one N = (dimension/resolution)$^2$ to one of information N = 3 × (number of structures). This allowed us to reduce the file size used for the master from a graphical one of 900MB to slightly in excess of 7MB for a 10cm diameter with an effective resolution of 1μm. A microscope picture of the grid taken at different positions across the grid (Fig.2) shows the variation of square sizes at different positions. Careful examination of the smaller structures will show variation within the single inset.

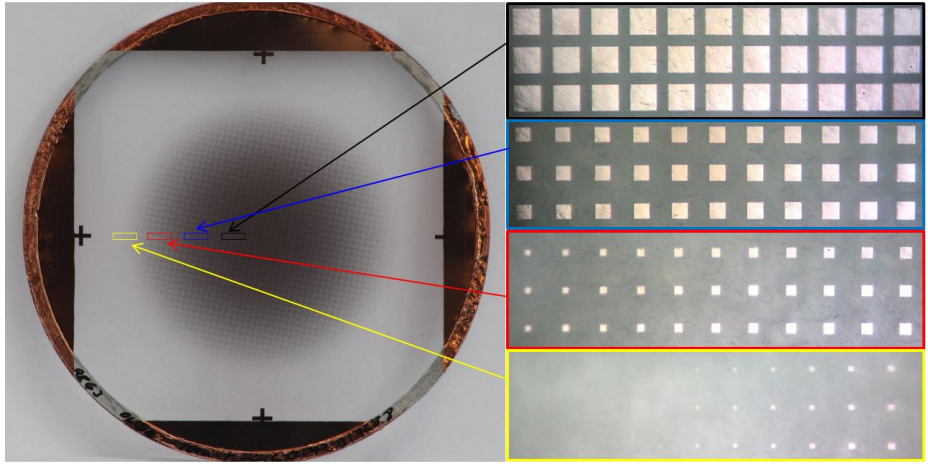

Fig.2. Composite figure with different parts of the single grid-mesh showing the variation of square size and filling factor with position.

The first step was to verify the replication of single meshes used as the building blocks for our GrIn lens. Spectral measurements were made at different positions (of the grid) and compared with the well established mesh theory [2,7]. A direct spectrum ratio with/without the lithographed structure performed with a Fourier Transform Spectrometer with a focused (but only moderately converging) beam provided preliminary information on the cut-off frequencies associated with different portions of the mesh. The effectiveness of the spectral filtering nature of the stack of capacitive meshes is shown in Fig.4.

*2.4 Grin lens prototype*

The hot-pressed [26] GrIn lens is a 10cm diameter disk of 2mm thickness as shown in Fig.3. Based on the gradient index equations Eq.2&3 detailed above, we designed the lens to focus radiation from a parallel beam to a point at a distance of $f \cong 25cm$ for a frequency of 150GHz ($\lambda = 2mm$). In order to verify the beam-converging properties of the Wood lens we used a Hg-arc lamp thermal source with a Band-Pass filter centred at 2mm to select the desired spectral portion of radiation. To detect the power at a well defined aperture in the lens focal plane we used a 4K cooled bolometric detector with a parabolic horn feed inside the dewar vacuum vessel and coupled this to external radiation via a back-to-back smooth-walled horn pair outside which created a beam waist at the dewar window. The front horn of this external pair was then stopped down to a 5mm aperture to probe the focal plane power. This arrangement overcomes the difficulties associated with the detector horn location inside the vacuum vessel allowing the focal plane to be sampled cleanly.

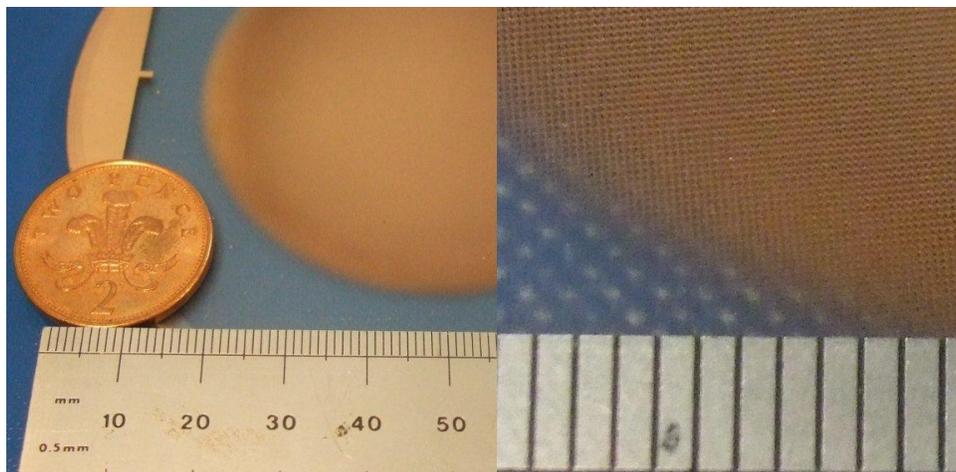

Fig.3. Picture of the GrIn lens formed from the hot-pressed grids. Detailed structure is difficult to observe not only due to the small structure size but also due to the alignment of subsequent grids.

To generate a plane parallel beam a 7cm diameter HDPE lens with a focal length of f=175mm was placed with its focus at the source image position of the Fourier Transform Spectrometer output. A second identical lens was placed at a distance of $\cong 400mm$ from the first to converge the beam on to the detector external horn aperture. A spectrum was taken in this configuration to provide the reference spectral throughput measurement which accounts for the spectral variations of all the optical components. The GrIn lens was subsequently inserted at the same position of the second re-focusing HDPE lens and a second spectrum was taken to compare the overall optical efficiency of the lens and its spectral characteristics. The resulting spectral ratio can be seen in Fig.4.

To show how the lens is performing across its aperture another spectrum was recorded with the central area obscured by a metal disc positioned on the parallel-beam side of the GrIn lens to provide the equivalent spectral ratio of only the external corona (2cm < r < 3.5cm). This data shows that the central part of the lens is not contributing to intensity at frequencies higher than ~ 420 GHz which is evidence of its behavior as a low-pass filter in accordance with the filtering properties of a stack of capacitive meshes [2,26].

Comparing the efficiency of the GrIn lens to that of the polyethylene one in Fig.4 two effects must be considered. Firstly at frequencies below the central cut-off (~ 350GHz) the efficiency can be calculated by integrating the reflection losses as a function of the lens radius, whereas for the polyethylene lens "n" is constant. The ratio of these two yields $\cong 0.75$ which is higher than the value measured in Fig.4. This is largely explained by the differing focal length of these lenses resulting in an over sized image at the detector aperture.

An ideal comparison would have required a thicker GrIn lens design which for the practical reasons discussed earlier was not implemented. Secondly for frequencies above the cut-off, where the lens is not expected to be used, the calculation of the reflection losses becomes more complex as it now also depends on the cut-off frequency profile.

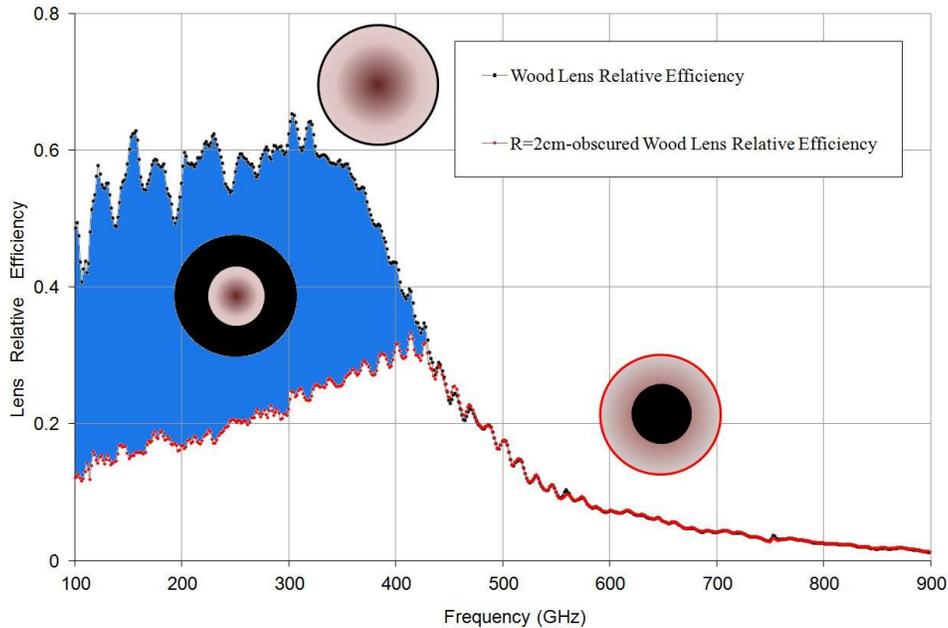

Fig.4. The ratio of spectral transmission data of the Grin Lens compared to the PE lens is shown in black dots. The same measurement performed with a central obscuration shows identical transmission above 420 GHz proving that at higher frequencies there is no transmission within the radius of the obscuration, and also that at these frequencies, the lens is perfectly functional. The lower efficiency at lower frequencies where no cut-off is present (of the red curve compared to the black) can be explained with the ratio of geometrical area corona remaining from the central obscuration. The plateau value of efficiency reached is lower than the optimal efficiency of the black curve in the same region for a de-focused position of the lens due to a longer focal distance.

A second experiment used the same HDPE lenses with a free standing chopped source (no FTS) and optical filters to define the spectral band of the now phase-locked detector. Using this we were able to measured the beam intensity near the axis by scanning the detector across the optical axis for three configurations: 1) no refocusing lens, 2) with an identical achromatic HDPE lens and 3) with our flat Wood lens (substituted as before for the final HDPE lens). The scan with no condensing optics (1) showed that the wavefront was uniform over a distance comparable to the lens aperture. The normalized beam profiles measured across the optical axis can be seen for the latter two cases in Fig.5. Deconvolving for the detecting horn aperture, we obtain values of 8.5mm and 6.8mm for the point spread function FWHM of the GrIn and the HDPE lenses respectively.

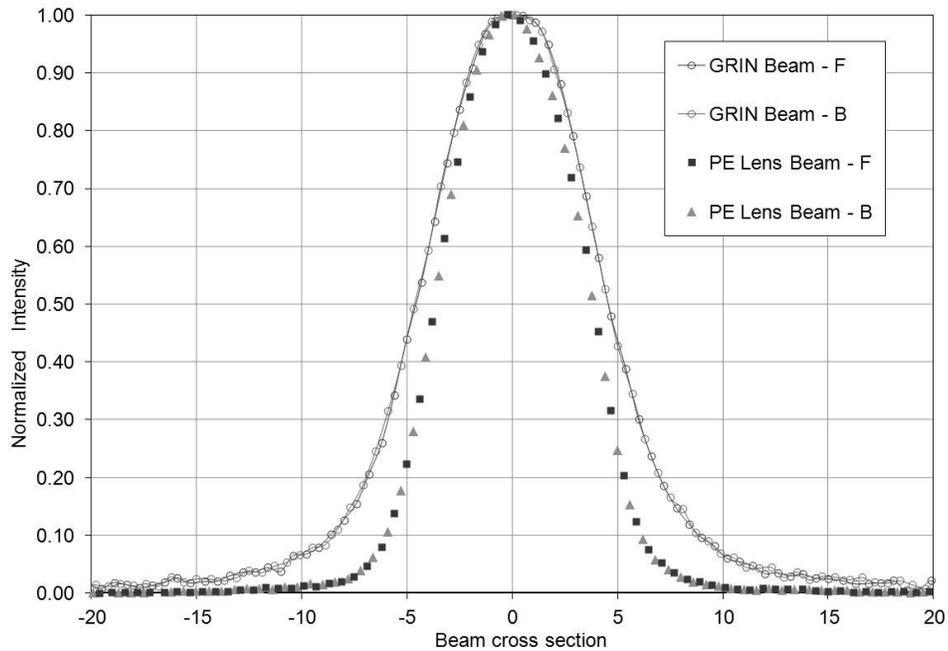

Fig.5. Normalized comparison of lens beam profiles. (Square and Triangle graphs): Cross-scan of canonical polyethylene lens. (Circle-line graph): Cross-scan of Gradient Index lens at the same position. The relative beam-width measured is larger in the GrIn lens (9.2mm) compared to the (7.5mm) of the PE lens due to the longer focal length. The red and blue points indicate separate scan averages in opposite directions (F=Left to Right, B=Right-to-Left).

## 3. Conclusions

We have shown how it is possible to build a 3D engineered material with a spatially graded refractive index normal to the surface by using stacked layers of 2D photo-lithographic patterned metal-meshes embedded in polypropylene. This design allows wide band coverage and has an optical efficiency currently limited by the impedance mis-match of the material equivalent index. This will be remedied in the future by effectively applying a graded refractive index anti-reflection coating in the direction of the optical axis (as achieved in Zhang09 for a uniform structure) to produce a fully efficient thin flat lens. The inherent filtering properties will prove invaluable to those designing photometric instruments where spectral filtering and beam control are now combined in a single optical element. The lens performs close to expectations with its properties largely determined by the metal mesh parameters. This work shows for the first time the combined power of spectral filtering and beam focusing in a single thin macroscopic device and opens a vast range of possibilities for future applications based on transformation optics.

**Acknowledgements**

We acknowledge the support of the Science Technology and Facilities Council with grant STFC No. ST/J001449/1.